# On the drift wave eigenmode crossing zero frequency in Tokamak

Z. Y. Liu[1†], Y. Z. Zhang[2], S. M. Mahajan[3], T. Xie[4]

[1]Institute for Fusion Theory and Simulation and Department of Physics,

Zhejiang University, Hangzhou, Zhejiang 310027, China

[2]Center for Magnetic Fusion Theory, Chinese Academy of Sciences, Hefei, Anhui 230026, China

[3]Institute for Fusion Studies, University of Texas at Austin, Austin, Texas 78712, USA

[4]School of Science, Sichuan University of Science and Engineering, Zigong, Sichuan 643000, China

## Abstract

The conventional ion temperature gradient (ITG) or $\eta_i$ mode is known to propagate in the ion diamagnetic direction. The mode frequency, however, may change in magnitude, even in sign, as we vary $\eta_i$. Investigation of a generic drift fluid model with warm ions ($T_i \approx T_e$) and adiabatic electrons, reveals that as $\eta_i$ decreases, the propagation characteristics of the unstable mode may change drastically – the mode frequency first decreases in magnitude, and reaches zero for a critical $\eta_i$. But as $\eta_i$ goes down further, the mode begins to propagate in the electron diamagnetic direction. The lower toroidal mode number perturbations are more prone to reversal in propagation direction. Even for $\eta_i = 0$, the mode remains unstable, drawing free energy form the density gradients. Since finite ion temperature appears to be essential for propagation in the electron direction, it is appropriate to introduce new terminology and call this wave the warm ion electron drift (WIED) mode. It is to be distinguished from the well-known dissipative electron drift wave that does survive in the cold ion limit, and will be, henceforth, called the cold ion electron drift (CIED) mode; it requires dissipation to be unstable. The model drift-wave system is explored within the framework of the two dimensional (2D) weakly asymmetric ballooning theory (WABT) for *local* eigenmode satisfying *natural boundary conditions*. The physics behind the excitation of the eigenmode crossing zero frequency is identified to be the reactive instability induced by the curvature coupling between the positive energy wave $\widehat{\omega}_+$ and the negative energy wave $\widehat{\omega}_-$, schematically represented as: $(\omega - \widehat{\omega}_+)(\omega - \widehat{\omega}_-) = \hat{\kappa}$, *i.e.*, a damped mode in electron direction coupled to a growing mode in the ion direction in non-dissipative slab limit. Apart from its intrinsic scientific value, this mechanism may shed some light onto the nature of tokamak (trapped-particle-free) edge turbulence observed in frequencies moderately lower than the electron diamagnetic frequency; understanding this phenomenon could be helpful in conceptual design around the edge region of future tokamak.

---

[†] Corresponding author. E-mail: lzy0928@mail.ustc.edu.cn

## 1. Introduction

A search into the origin and nature of the drift wave turbulences at the tokamak edge attracted considerable attention some thirty years ago [1]. To appreciate how difficult it was to identify the nature of the modes supporting edge turbulences observed in experiments like those on TEXT [1], one simply has to remember that even exotic driving mechanisms (atomic processes like ionization, charge exchange, radiation and impurity [2, 3]), far beyond the conventional thermodynamic free energy source like gradients, were also invoked. In hindsight, there are two factors mostly responsible for the reigning confusion:

(1) Most theory-simulation work was, then, limited to slab models [4, 5] in which all dissipative modes propagating in electron diamagnetic direction suffer from strong (magnetic) shear stabilization. Their exclusion as well as preclusion of the well-known collisionless trapped electron mode (TEM) (because of high collisional de-trapping at the low temperature edge), left little choice with the theorists.

(2) The experiment diagnostic tools were not advanced enough to allow a clean separation between the intrinsic mode frequency (in the plasma rest frame) and the Doppler shift induced by plasma rotation. There was further complication due to the prevailing belief that the Doppler shift would be reduced well below the diamagnetic frequency via the strong flow-damping from magnetic pumping [6].

Thanks to positive developments in both theory and experiment, the current situation is drastically different. Nowadays, the experimentalists are able to separate plasma rotation from the observed 'mode frequency'. In Ref. [7], for example, we see the broadened dispersion coarse 'curve' obtained near zero frequency in rotating frame; it pertains to turbulence near the plasma edge for $\rho = 0.98$ on ASDEX-U. The simulation reported in [8] (that keeps the magnetic curvature by making use of flux tube coordinate) attributes the turbulence to an interchange mode. In fact, the experiments seem to identify a host of modes in electron diamagnetic direction: the quasi-coherent mode (QCM) [9, 10], the edge coherent mode (ECM) [11, 12], and the weak coherent mode (WCM) [13-16]. The origin of the mode instability was attributed either to TEM or to interchange drive.

It is well known that the toroidal coupling greatly mitigates shear stabilization; it was shown, for example, by the 1D ballooning theory for drift wave in the $\delta_e$- model [17]. The 1D ballooning theory, however, has only limited usefulness; it may provide a good approximation to the 'local eigenvalue' (containing an unknown parameter $\lambda$, see Equation (B3) and the definition for terms 'local' and 'global' in Appendix B), but a 'global eigenvalue' cannot be determined without the simultaneous knowledge of the mode structure; the latter is also required for the initial data that must be fed into an iterative approach, such as shifted inverse power method (SIPM) [18], for solving eigenmode problems. Fortunately, there does exist a 2D ballooning theory, the WABT [19] that can supply the needed information for the purely numerical SIPM [20]. WABT, along with SIPM, has been successfully applied to analyze kinetic ITG [21], TEM [22]. The authors have demonstrated its usefulness by calculating Reynolds' stress and group velocity in real space for investigating, for example, the nonlinear generation of low frequency zonal flow and intermittent excitation of geodesic acoustic mode (GAM) [23-26].

It is to be emphasized that the eigenvalue problem in WABT is solved by applying the "natural boundary conditions", in contrast to what is done in most existing simulations, *e.g.*, [27-29]; these employ Dirichlet-Neumann boundary condition for their global calculations. We use the term 'natural boundary condition' both as descriptive and distinguishing: we seek the entire solution that matches the natural form of the exact solution in the asymptotic region - the region near the coordinate boundaries. This asymptotic approach is to be distinguished from the one based on Dirichlet-Neumann boundary conditions where the eigenvalue is obtained by, somewhat artificially, forcing the function (or its derivative) to assume some specific pre-assigned value at a point. It is interesting to note that, perhaps the first major paper on drift wave eigenmode (called universal instability those days) invoked, precisely the natural - outgoing wave boundary conditions - to solve the 1D drift wave in the sheared slab model [30].

The WABT solution, as the solution of the natural boundary value problem, is a local toroidal mode (not confused with the definition of 'local' and 'global' used for eigenvalue in Appendix B); it pertains to each rational surface across a radial domain for a given toroidal mode number $n$. The natural boundary value problem is implemented in the Zhang-Mahajan 2D ballooning representation [31] in the $(x, l)$ representation. The $(x, l)$ representation is chosen for explicitly displaying translational invariance [32]. In the corresponding Fourier-ballooning space, the variable $k$ is the Fourier conjugate of $x - l$ while the Floquet phase $\lambda$ is the Fourier conjugate of $l$. In the leading order system, all dependence comes through $x - l$ (translational invariant). It is, however, the translational symmetry breaking (TSB) terms (appearing as $x$ and $l$) that provide the rich structure to this system; such terms including the sideband contributions $\sim l/n$, and slow radial variation of equilibrium profile $\sim x/n$, are assigned to a higher order for large $n$. Neglecting all TSB terms reduces the 2D problem to the conventional 1D ballooning equation. In contrast to other ballooning theories where the translational invariance is not explicitly exploited, the higher order equation of WABT - the equation for Floquet phase distribution (FPD) - is built by collecting all TSB terms to the second order. As suggested by Lee-Van Dam theory [33], the asymptotic solutions of the FPD equation should be evanescent; the WABT, therefore, automatically leads to solution satisfying natural boundary condition required for local mode. The results from the WABT solution could be further refined by SIPM [18], This step is not merely a quantitative refinement; it removes the restrictions required by the asymptotic conditions of WABT [34] since the convergence of SIPM is fully determined by the equation itself, and independent of those asymptotic conditions.

The main theme of this paper is to investigate the phenomenology and physics of propagation reversal. It occurs, i.e., the mode begins to propagate in the electron direction, when both $\eta_i$ (the ratio of density to ion temperature gradient length) and the mode number *n* are in the smaller range. For larger $\eta_i$ and *n*, the mode propagates in the ion direction as shown in Figure 3. The ITG and WIED are shown to be the same mode (owing to smooth transition at zero frequency) that exists possible only for finite temperature ions; it is legitimate, then, to coin a unified name for the synthetic mode - the warm ion drift (WID) mode. The WID mode is explored in Section 6 in detail. It is essentially associated with the warm ion branch, one of two branches in the slab limit, which does not have cold ion limit and is the negative energy wave destabilized by magnetic shear. It is coupled to the positive energy wave CIED mode in the other (cold ion) branch via magnetic curvature, and analytically continued to full coupling from slab configuration.

The paper is organized as follows. The generic drift fluid model and the resulting 2D eigenmode equation in $(x, l)$ representation, are described in Section 2. In Section 3, three theoretical and computation approaches are presented, namely, 1D ballooning theory (referred to be 1D-BT henceforth), WABT, and SIPM. In Sections 4 and 5, the basic findings of this work on the solution of the eigenvalue problem - eigenvalues and associated mode structure - are discussed in detail. Particular attention is paid to the region of propagation reversal. The physics behind propagation reversal is discussed in Section 6, while Section 7 is devoted to a summary and conclusions. The paper is supplemented by several appendices. In Appendix A all terms of the generic drift fluid model in $(x, l)$ representation, are explicitly derived up to the second order in the expansion parameter. The detailed procedures of WABT and SIPM are reviewed, respectively, in Appendix B and C. In Appendix D, 1D-BT is solved in the slab model where the warm and cold ion branches are defined clearly.

## 2. The generic drift fluid model in $(x, l)$ representation

The generic drift fluid model of this paper is built on a large-aspect-ratio, up-down symmetric tokamak equilibrium with concentric circular magnetic surfaces [24]. The distinguish features of this model are (a) electron response is adiabatic; (b) it is trapped-particle free, and (c) it is suitable for edge with high collision. We start from the three linear moment equations - the continuity equation of warm ion, ion parallel momentum equation, and pressure evolution equation

$$\frac{\partial}{\partial t}\left(1-\rho_s^2\nabla_\perp^2\right)\hat\varphi = -\left(1+\bar\eta_i\rho_s^2\nabla_\perp^2\right)\boldsymbol{v}_*\cdot\nabla\hat\varphi - \nabla_\parallel\tilde v_{i\parallel} + \boldsymbol{v}_D\cdot\nabla(\hat\varphi+\hat p), \tag{1}$$

$$\frac{\partial \tilde v_{i\parallel}}{\partial t} = -c_s^2\nabla_\parallel(\hat\varphi+\hat p), \tag{2}$$

$$\frac{\partial \hat p}{\partial t} = -\bar\eta_i\boldsymbol{v}_*\cdot\nabla\hat\varphi, \tag{3}$$

where the quantities with tilde (caret) stand for perturbed (normalized) ones, $\hat\varphi \equiv e\tilde\varphi/T_{e0}$, $\hat p \equiv \tilde p_i/p_{e0}$, those with subscript zero for equilibrium. The velocities $\boldsymbol{v}_* \equiv -\rho_s c_s \boldsymbol{b}\times\nabla\ln n_0$ and $\boldsymbol{v}_D \equiv 2\rho_s c_s \boldsymbol{\kappa}\times\boldsymbol{b}$ are due to diamagnetic and curvature drift respectively, $\rho_s \equiv \sqrt{m_i T_{e0}}/eB$ is the ion Larmor radius at electron temperature, $c_s \equiv \sqrt{T_{e0}/m_i}$ is the ion sound speed, $\boldsymbol{b}$ is the unit vector along the equilibrium magnetic field, $\boldsymbol{\kappa}\equiv \boldsymbol{b}\cdot\nabla\boldsymbol{b}$ is the magnetic curvature, and $e$ is the unit electric charge. $\bar\eta_i \equiv \tau_i(1+\eta_i)$, $\tau_i \equiv T_{i0}/T_{e0}$, $\eta_i \equiv d\ln T_{i0}/d\ln n_0$ is the ratio of ion temperature gradient to density gradient, which is essential to the ITG instability.

The system (1)-(3) leads to the eigenmode equation [20]

$$\left(\rho_s^2\nabla_\perp^2 - \frac{c_s^2}{\omega^2}\nabla_\parallel^2 - \frac{\omega+i\boldsymbol{v}_*\cdot\nabla}{\omega-i\bar\eta_i\boldsymbol{v}_*\cdot\nabla} + \frac{i\boldsymbol{v}_D\cdot\nabla}{\omega}\right)\hat\varphi = 0. \tag{4}$$

In the toroidal coordinates $(r,\vartheta,\zeta)$, the 2D mode can be expressed in the $(x,l)$ representation near the rational surface $r_j$, defined by $m_j \equiv nq(r_j)$ for integer $m_j$ and given q-profile. The subscript $j$ will be suppressed in $m_j$ henceforth, because in this paper we are dealing solutions at only one rational surface $r_j$. The $(x,l)$ representation is defined by

$$\hat\varphi(r,\vartheta,\zeta) = \exp[i(n\zeta-m\vartheta)]\sum_l \varphi_l(x)\exp(-il\vartheta), \tag{5}$$

where $x \equiv k_\vartheta \hat s (r-r_j)$, as the new radial variable, $k_\vartheta \equiv m/r_j$ and $\hat s \equiv d\ln q(r_j)/d\ln r$ is the magnetic shear, $n$ is the toroidal mode number (a good quantum number in this paper), m is $m_j$ with subscript $j$ suppressed. The rational surface $r_j$ is called central rational surface of the local toroidal mode, uniquely defined by the pair of integers $(n, m)$; for a monotonic $q(r)$, it is distinct from other rational surfaces (sidebands) contained in the local mode. Conversion of Equation (4) into $(x,l)$ representation (to second order) is carried out in Appendix A, and reads

$$\left[\hat k_\vartheta^2 \hat s^2 \frac{d^2}{dx^2} - \hat k_\vartheta^2 + (x-l)^2\frac{\hat\omega_s^2}{\hat\omega^2} - \frac{\hat\omega-1}{\hat\omega+\bar\eta_{i,j}}\right]\hat\varphi_l - \frac{1}{2}\frac{\hat\omega_D}{\hat\omega}\left[\hat s\frac{d}{dx}(\hat\varphi_{l+1}-\hat\varphi_{l-1})+(\hat\varphi_{l+1}+\hat\varphi_{l-1})\right]$$

$$+\left\{-\frac{(\hat s-1)\hat\omega_D}{2\hat s\hat\omega}(\hat\varphi_{l+1}+\hat\varphi_{l-1}) + \left[A_1(\hat\omega)-2(x-l)^2\frac{\hat\omega_s^2}{\hat\omega^2}\right]\hat\varphi_l\right\}\left(\frac{l}{m}\right) \tag{6}$$

$$+\left\{\frac{(\hat s-1)\hat\omega_D}{2\hat s^2\hat\omega}(\hat\varphi_{l-1}+\hat\varphi_{l+1}) + \left[A_2(\hat\omega)+3(x-l)^2\frac{\hat\omega_s^2}{\hat\omega^2}\right]\hat\varphi_l\right\}\left(\frac{l}{m}\right)^2 = 0$$

where $\hat{k}_g \equiv \rho_{s,j} k_g$, $\hat{\omega} \equiv \omega / \omega_{*,j}$, $\hat{\omega}_s \equiv c_{s,j} / qR\omega_{*,j}$, $\hat{\omega}_D \equiv \omega_{D,j} / \omega_{*,j}$, subscript $j$ denotes equilibrium quantities at the rational surface $r_j$, $A_1(\hat{\omega})$ and $A_2(\hat{\omega})$ are given in Equation (A10) and (A11) respectively. The diamagnetic drift frequency $\omega_* \equiv -\rho_s c_s k_g / L_n$ and the curvature drift frequency $\omega_D \equiv -2\rho_s c_s k_g / R$, $R$ is the major radius, $L_n$, $L_{T_e}$ and $L_{T_i}$ are the density, electron and ion temperature gradient length respectively.

## 3. Theoretical and computational approach

For a high $n$ local mode pertaining to a given rational surface $r_j$, the 2D ballooning representation [31, 19, 20]

$$\varphi_l(x) = \frac{1}{2\pi} \int_{-\pi}^{\pi} d\lambda \int_{-\infty}^{+\infty} dk\, e^{ik(x-l)-i\lambda l} \varphi(k,\lambda), \tag{7}$$

will map 2D $(x, l)$ space to the 2D $(k, \lambda)$ space. A perturbation theory is developed by assuming $\varphi(k,\lambda) := \psi(\lambda)\chi(k,\lambda)$, where $\psi(\lambda)$ is a fast varying function in $\lambda$ (known as Floquet phase distribution (FPD)) while $\chi(k,\lambda)$ is the standard solution of ballooning equation in which $\lambda$ appears as a parameter. $\psi(\lambda)$ is localized around some $\lambda_*$. The Lee-Van Dam representation [33] is just the singular limit $\psi(\lambda) \to \delta(\lambda - \lambda_*)$ ($\delta$ denotes Dirac delta function). Notice that Equation (7) is a proper mathematical transform (a unique inverse transform exists).

In the 2D $(k, \lambda)$ space, Equation (6) is expressed as

$$\left[ L_0(k,\lambda;\hat{\omega}) + \frac{iL_1(k,\lambda;\hat{\omega})}{n} \frac{\partial}{\partial \lambda} + \frac{L_2(k,\lambda;\hat{\omega})}{n^2} \frac{\partial^2}{\partial \lambda^2} + \cdots - \Omega(\hat{\omega}) \right] \varphi(k,\lambda) = 0, \tag{8}$$

$$\Omega(\hat{\omega}) = -\frac{\hat{\omega}}{\hat{\omega}_s}\left( \hat{k}_g^2 + \frac{\hat{\omega}-1}{\hat{\omega}+\bar{\eta}_{i,j}} \right), \tag{9}$$

$$L_0(k,\lambda;\hat{\omega}) = \frac{\hat{\omega}_s}{\hat{\omega}} \frac{\partial^2}{\partial k^2} + \frac{\hat{\omega}}{\hat{\omega}_s} \hat{k}_g^2 \hat{s}^2 k^2 + \frac{\hat{\omega}_D}{\hat{\omega}_s}\left[ \cos(k+\lambda) + \hat{s} k \sin(k+\lambda) \right], \tag{10}$$

$$L_1(k,\lambda;\hat{\omega}) = \frac{1}{q_j} \frac{\hat{\omega}}{\hat{\omega}_s}\left[ A_1(\hat{\omega}) - \frac{(\hat{s}-1)\hat{\omega}_D}{\hat{s}\hat{\omega}} \cos(k+\lambda) + 2\frac{\hat{\omega}_s^2}{\hat{\omega}^2} \frac{\partial^2}{\partial k^2} \right], \tag{11}$$

$$L_2(k,\lambda;\hat{\omega}) = \frac{1}{q_j^2} \frac{\hat{\omega}}{\hat{\omega}_s}\left[ A_2(\hat{\omega}) + \frac{(\hat{s}-1)\hat{\omega}_D}{\hat{s}^2\hat{\omega}} \cos(k+\lambda) - 3\frac{\hat{\omega}_s^2}{\hat{\omega}^2} \frac{\partial^2}{\partial k^2} \right]. \tag{12}$$

### 3.1. One dimensional ballooning theory (1D-BT)

The structure of Equation (8) is clearly suggestive: in the large $n$ limit, $(\partial/\partial\lambda)/n$ is the effective measure of the symmetry breaking terms. One, thus, develops a successive perturbation theory in which, the leading order equation

$$\left[L_0(k,\lambda;\hat{\omega}) - \Omega(\hat{\omega})\right]\chi(k,\lambda) = 0, \tag{13}$$

is the traditional ballooning equation with the Floquet phase $\lambda$, appearing as a parameter; the phase $\lambda$ is equivalent to the so-called ballooning angle $\theta_0$, representing the center position of the mode in 'poloidal direction' in the Connor-Hastie-Taylor (CHT) representation [35]. When the mode is ballooned around the mid-plane on bad curvature side $(\lambda \approx 0)$, it seems reasonable to seek for the eigenvalue starting from $\lambda = 0$. As shown in Section 4, the eigenvalue found in one dimensional ballooning theory (1D-BT) (corresponding to $\lambda = 0$) is not far away from the 2D eigenvalue solution of WABT and SIPM.

Substituting Equation (10) into Equation (13), the ballooning equation at $\lambda = 0$ is

$$\left[\frac{\partial^2}{\partial k^2} + V(k) - \frac{\hat{\omega}}{\hat{\omega}_s}\Omega(\hat{\omega})\right]\chi(k) = 0, \tag{14}$$

just like the Schrödinger equation. The potential function is defined as

$$V(k) = \frac{\hat{\omega}^2}{\hat{\omega}_s^2}\hat{k}_g^2\hat{s}^2k^2 + \frac{\hat{\omega}\hat{\omega}_D}{\hat{\omega}_s^2}(\cos k + \hat{s}k\sin k). \tag{15}$$

At large $k$, Equation (14) becomes a Weber-Hermite equation allowing the asymptotic boundary condition for an outgoing wave [30]

$$\chi(k) \xrightarrow[|k|\to\infty]{} \exp\left(\frac{i}{2}\frac{\hat{\omega}}{\hat{\omega}_s}\left|\hat{k}_g\right|\hat{s}k^2\right). \tag{16}$$

By imposing this boundary condition, Equation (14) can be solved via a 1D shooting code.

### 3.2. Weakly asymmetric ballooning theory (WABT)

For the solution of entire 2D eigenvalue problem, we must construct and analyze a differential equation in the second dimension; *i.e*, in the phase $\lambda$. Such a system emerges in higher orders; the relevant asymptotic theory is made possible by two small parameters: (1) the first is what we encountered earlier $\varepsilon_B \equiv (1/n)(\partial/\partial\lambda) \ll 1$; this will allow us to truncate Equation (8) to second order; (2) the second is a little less obvious. For a mode peaked around mid-plane, but with weak up-down asymmetry, $\lambda_* \approx 0$, it emerges as $\Xi \equiv \bar{L}_1/2\bar{L}_2 \ll 1$ [20], where $\bar{L}_{1,2}$ are the average value of $L_{1,2}$ in Equation (8). When these two conditions are satisfied, the 2D system splits in a system of two 1D ordinary differential equations. The first is $\lambda$-parameterized ballooning equation, an ordinary differential equation in $k$

$$\left[L_0(k,\lambda;\hat{\omega}) - \Omega(\lambda)\right]\chi(k,\lambda) = 0, \tag{17}$$

that will yield a $\lambda$ dependent eigenvalue. This will serve as an input to the next order differential equation in $\lambda$ obtaining by averaging over the ballooning variable $k$. The second equation, which we call the FPD equation,

$$\frac{d^2\psi(\lambda)}{d\lambda^2} + P(\lambda)\frac{d\psi(\lambda)}{d\lambda} + Q(\lambda)\psi(\lambda) = 0, \tag{18}$$

with

$$P(\lambda) \equiv \frac{n}{\overline{L}_2^{(0)}(\lambda;\hat{\omega})}\left[i\overline{L}_1^{(0)}(\lambda;\hat{\omega}) + \frac{2\overline{L}_2^{(1)}(\lambda;\hat{\omega})}{n}\right], \tag{19}$$

$$Q(\lambda) \equiv \frac{n^2}{\overline{L}_2^{(0)}(\lambda;\hat{\omega})}\left[\Omega(\lambda) - \Omega(\hat{\omega}) + \frac{i\overline{L}_1^{(1)}(\lambda;\hat{\omega})}{n} + \frac{\overline{L}_2^{(2)}(\lambda;\hat{\omega})}{n^2}\right], \tag{20}$$

$$\overline{L}_s^{(j)}(\lambda;\hat{\omega}) \equiv \frac{\int_{-\infty}^{\infty} dk\, \chi^*(k,\lambda) L_s(k,\lambda;\hat{\omega})\frac{\partial^j \chi(k,\lambda)}{\partial \lambda^j}}{\int_{-\infty}^{\infty} dk\, \chi^*(k,\lambda)\chi(k,\lambda)} \quad (s=1,2,\ j=1,2), \tag{21}$$

is rather complicated but is just an O.D.E in $\lambda$. We solve the two coupled ordinary differential equations by an iterative shooting code, which is described in Appendix B in detail.

### 3.3. Shifted inverse power method (SIPM)

SIPM [18] is an iterative approach for eigenvalue problem on grids of finite difference. It is powerful and relatively simple as long as the initial guess for both eigenvalue and mode structure is good enough. The only prerequisite of SIPM is the equation and the initial guess without invoking asymptotic theory, in contrast to WABT which is the 2D (asymptotic) ballooning theory, although presented numerically. In this paper SIPM is implemented in the $(x, l)$ representation. The initial guess for $\omega$ and $\varphi_l(x)$ can be obtained by making use of the WABT solution with mode structure $\varphi(k,\lambda) = \chi(k,\lambda)\psi(\lambda)$ via Equation (7). The iterative procedure of SIPM is given in Appendix C.

### 4. Numerical results for eigenvalue

For purpose of illustration the following basic equilibrium parameters are chosen at $\rho_j \equiv r_j/a = 0.6$ of L-mode discharge on HL-2A [36], $R = 1.65$m, $a = 0.4$m, $B = 1.35$T, $T_e = 250$eV, $L_{T_e} = 5.5$cm, $q = 1.7$, $\hat{s} = 0.9$, $L_n/R = 0.1$, $\tau_i = 1$. Toroidal mode number $n$ and $\eta_i \equiv L_n/L_{T_i}$ are variables. The variable $n$ may result in the radial position error, because for the fixed $\rho_j = 0.6$ the corresponding poloidal number defined by $m_j \equiv nq(r_j)$ may not be an integer. The round-off integer is then chosen for a new radial position closest to $\rho_j = 0.6$. However, for high $n$ this error is small and negligible. The subscript $j$ denoting the specific rational surface will also be neglected in the following presentation, because in this paper only one rational surface is discussed.

The eigenvalue solutions of 1D-BT, WABT and SIPM, for different toroidal numbers $n$ and $\eta_i = 3.0$, are displayed in Figure 1. One can see that eigenvalue of WABT is almost the same as that of SIPM, implying WABT indeed provides a good initial guess for SIPM. Eigenvalue of 1D-BT is close to both, however, with a little higher real frequency and lower growth rate. The most interesting result is that as toroidal number $|n|$ decreases (still remaining large), the real frequency of ITG in ion diamagnetic direction ($\omega < 0$), reduces towards zero, even falls into electron diamagnetic direction ($\omega > 0$). The propagation direction reversal occurs around $n = -35$ for WABT and SIPM, however, at a little higher $|n|$ for 1D-BT. It means that the 2D effects tend to impede propagation reversal with deceasing $n$.

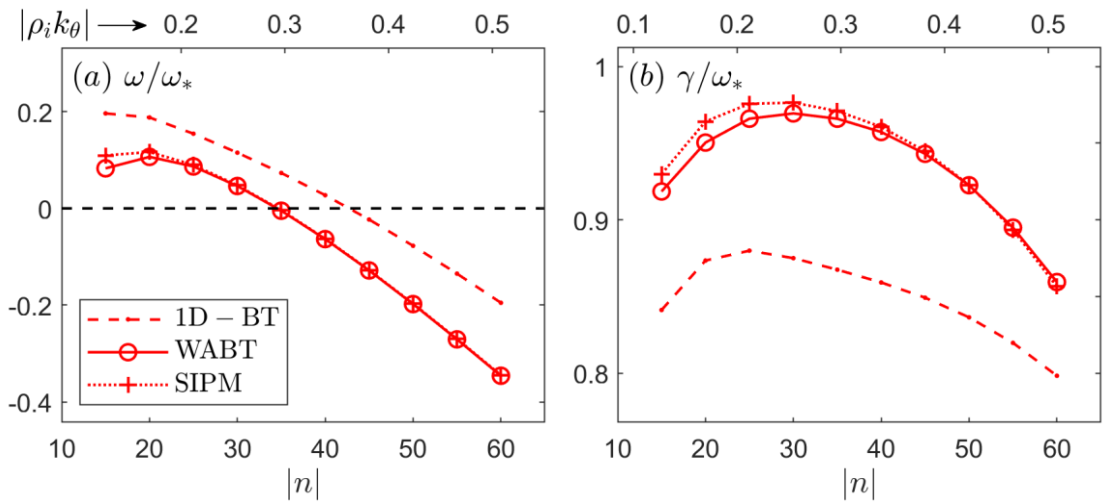

Figure 1. (a) Normalized real frequency $\omega/\omega_*$ and (b) growth rate $\gamma/\omega_*$ versus toroidal mode number $|n|$ for $\eta_i = 3.0$. Dashed-dot line: 1D-BT, solid-circle line: WABT, dotted-plus line: SIPM.

Recall that $\omega_*$ is positive since $k_\vartheta$ is negative

In Figure 2, the plots display the eigenvalue solutions versus $\eta_i \equiv L_n/L_{T_i}$ (with fixed $L_n$) for two different toroidal numbers, $n = -20$ and $n = -50$. There is propagation reversal for $n = -50$, however, for $n = -20$ propagation is always in electron direction even $\eta_i$ goes up to 5. One may compare the eigenvalues at $\eta_i = 3$ with those in Figure 1. The growth rate increases with increasing $\eta_i$, consistent with the consensus that ITG mode is driven by $\eta_i$. However, as $\eta_i$ approaches zero, there is still finite growth rate. The mode for very small $\eta_i$ is actually the WIED mode. One can readily see that the mode with $n = -20$ is always in electron direction regardless the value of $\eta_i$, whereas the mode with $n = -50$ alters direction of propagation per increasing $\eta_i$. In general, the lower toroidal number $n$ and lower $\eta_i$ favor mode frequency in the electron direction. This behavior could be seen more clearly in Figure 3 in which the borderline curves (for propagation reversal) are drawn in the $\eta_i$-$n$ parameter space.

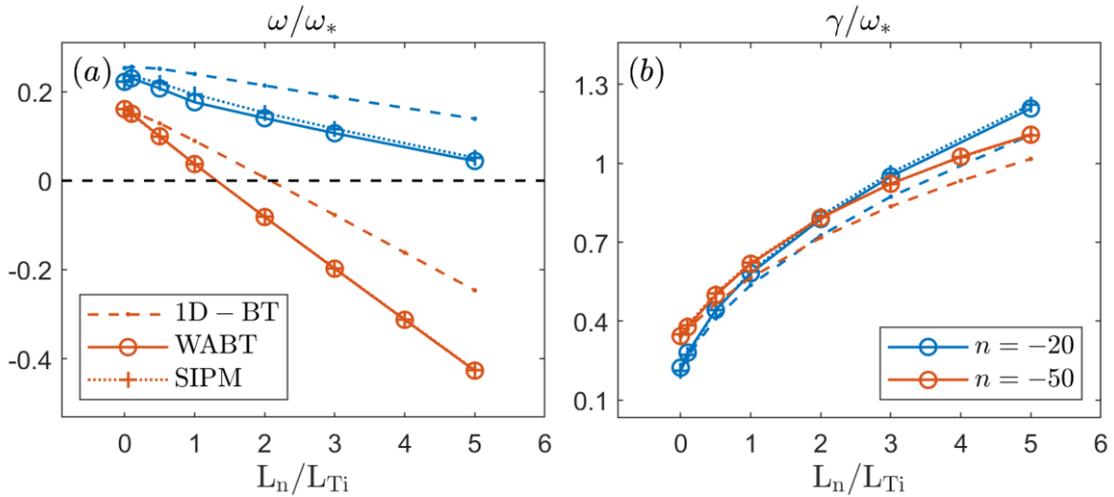

Figure 2. (a) Normalized real frequency $\omega/\omega_*$ and (b) growth rate $\gamma/\omega_*$ versus $\eta_i \equiv L_n/L_{T_i}$.

Dashed-dot line: 1D-BT, solid-circle line: WABT, dotted-plus line: SIPM.

Blue (Red) line for $n = -20, -50$ respectively.

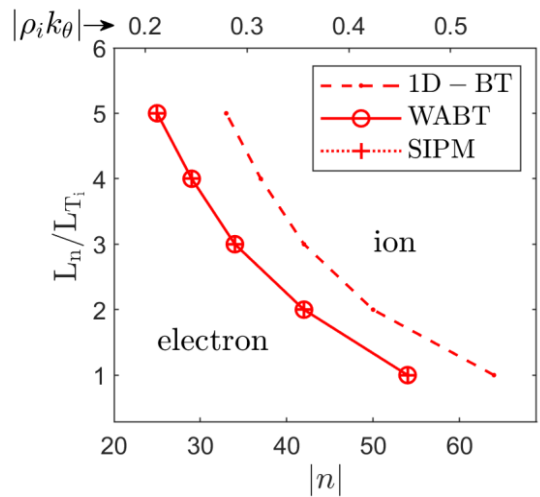

Figure 3. Borderline of propagation direction reversal in $\eta_i$-$n$ parameter space.

Dashed-dot line: 1D-BT, solid-circle line: WABT, dotted-plus line: SIPM.

The reasons for the mode frequency reversal are delineated in Section 6.

## 5. Numerical results for mode structure

In this section we will compare the eigen structures for the opposite extremities of the mode, WIED (for smaller $\eta_i$ and $n$) and ITG (for larger $\eta_i$ and $n$), typically case A: for $n = -20$, $\eta_i = 1.0$ and case B: for $n = -50$, $\eta_i = 5.0$.

## 5.1. The solution from ballooning theory

The effective potential of the ballooning equation, and the corresponding ballooning wave function are displayed in Figure 4 for both A and B. It seems that the 2D effects cause little modification to the 1D-BT mode structure. The width of ballooning wave function of WIED, however, is much wider than that of the ITG. The FPD from WABT, shown in Figure 5, is vastly different for A and B. For the former FPD looks like monopole-dipole for real and imaginary part, while it looks like quadrupole-quadrupole for the latter.

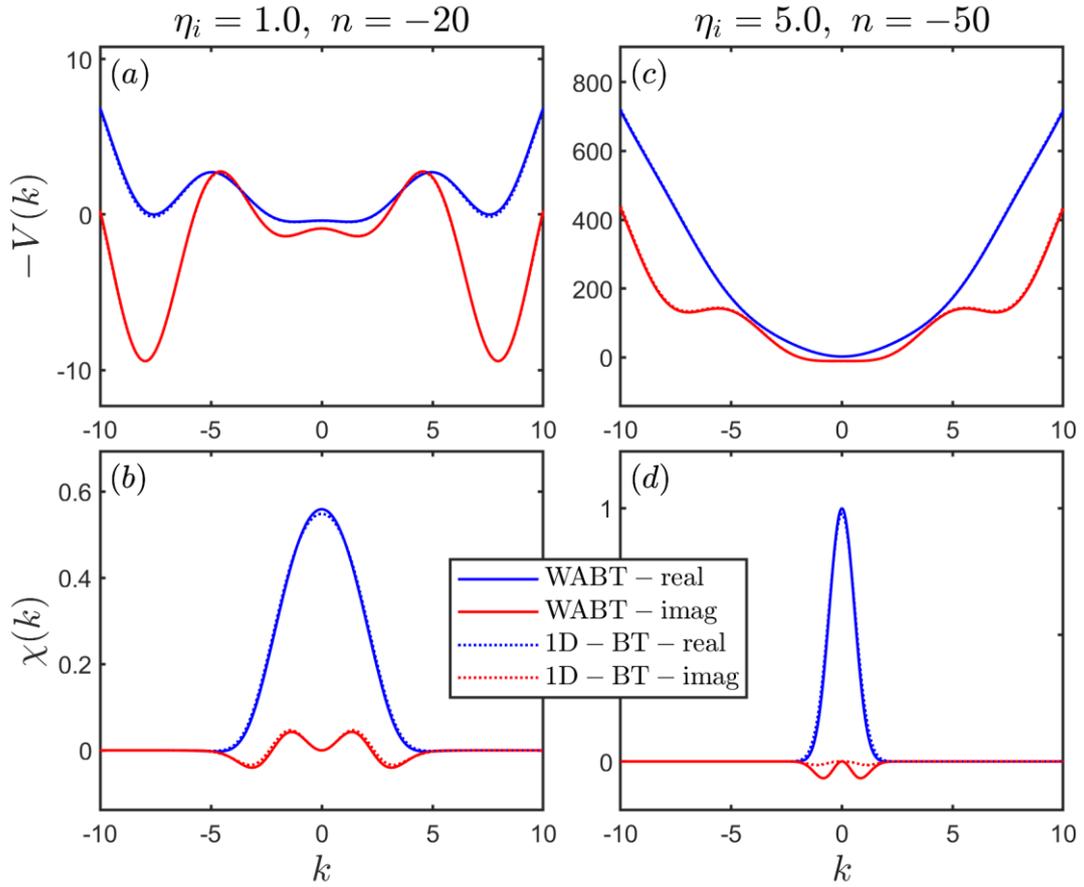

Figure 4. The potential function defined in Equation (15) and ballooning wave function from WABT and 1D-BT, (a/b) for case A, (c/d) for case B

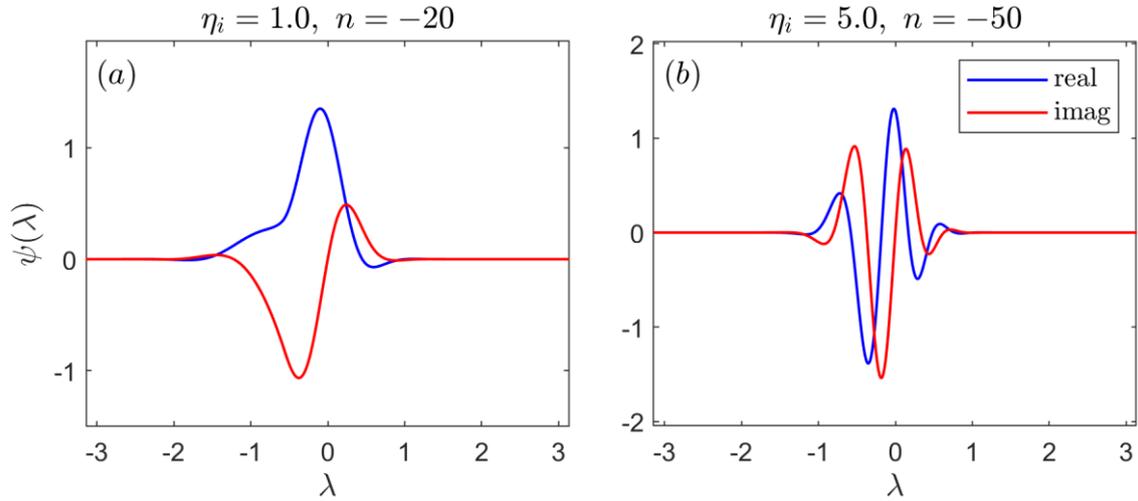

Figure 5. FPD in WABT, (a) for case A, (b) for case B

### 5.2. The mode structure in $(x, l)$ representation

The 2D wave function in $(x, l)$ representation is displayed in Figures 6A and 6B from both the WABT and SIPM. The modification from SIPM is visible, however, not significant. One can see a huge radial shift from the central rational surface in case B. The number of sidebands (including the central rational surface) contributing to the mode (when measured by the radial half width) is 6. Cases A (from 0 to 5) and B (from 6 to 11) are similar as shown in the SIPM solution. The radial extension for both A and B, measured in $x$, is ~5. These two values are not sensitive to toroidal number $n$, which could be attributed to the $n$-independence of curvature term in the ballooning equation.

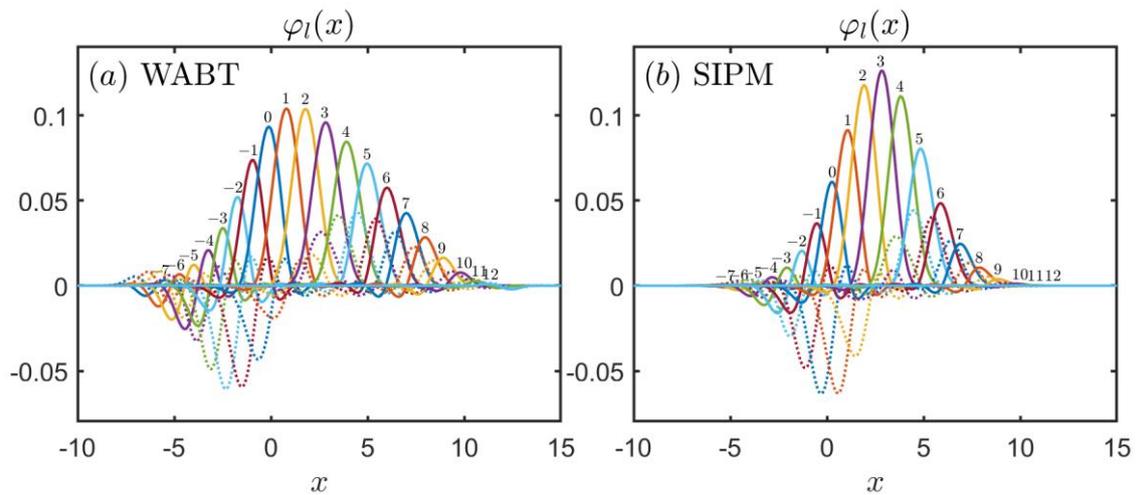

Figure 6A. Wave functions $\varphi_l(x)$ for different poloidal sideband number $l = -7, -6, \ldots, 11, 12$, $n = -20$, $\eta_i = 1.0$. (a) WABT, (b) SIPM. Solid/dotted line: real /imaginary part respectively.

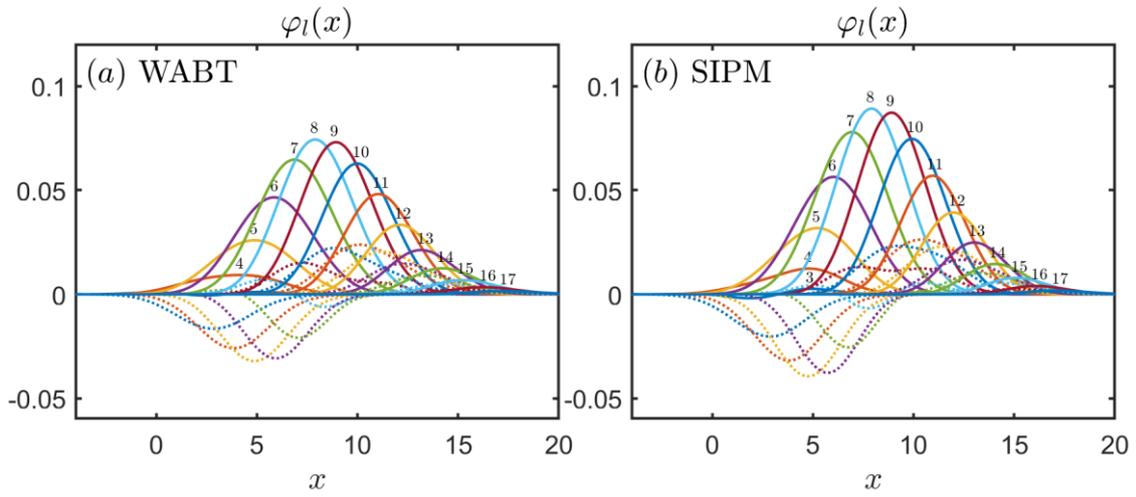

Figure 6B. Wave functions $\varphi_l(x)$ for different poloidal sideband number $l = 3, 4, \ldots, 16, 17$, $n = -50$, $\eta_i = 5.0$. (a) WABT, (b) SIPM. Solid/dotted line: real /imaginary part respectively.

### 5.3. The 2D mode structure on tokamak cross section

In Figures 7A and 7B, the 2D graphics for the mode structure, as the contour plot of $\text{Re}\varphi(r, \vartheta)$, are displayed for both WABT and SIPM. Two characteristics are found consistent with those observed in Figures 6A and 6B: (1) ITG has a significant radial shift from the central rational surface. Such a kind of strong radial shift may have severe implication to experiment for the equilibrium parameter associated with the mode observed, and (2) radial width of WIED mode is fairly broader than that of ITG.

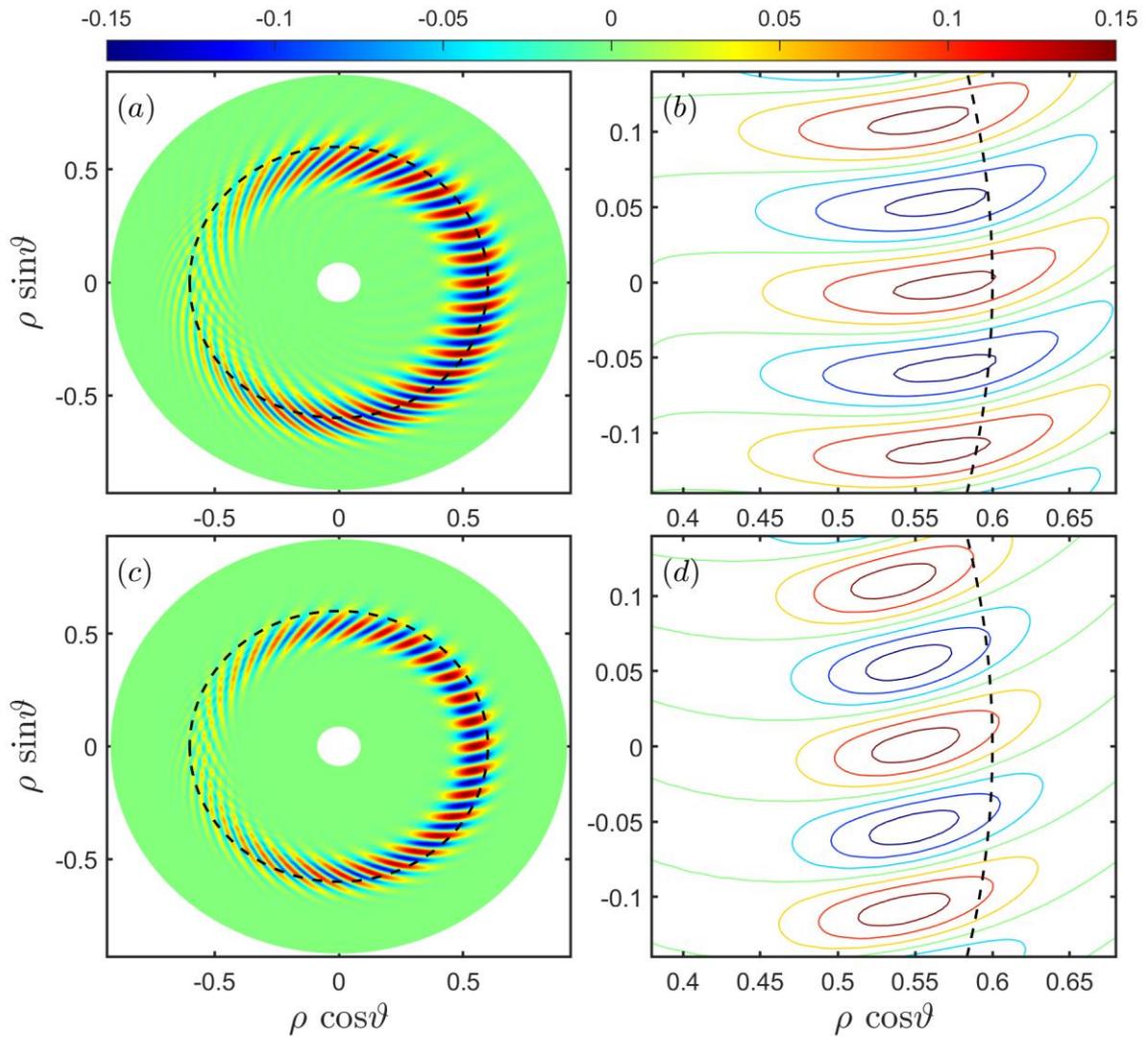

Figure 7A. Real part of the two dimensional wave function $\text{Re}\varphi(r,\vartheta)$, $n = -20$, $\eta_i = 1.0$.

(a) WABT, (c) SIPM. (b/d): close-up in bad curvature regime of (a/c) respectively.

The black dashed line represents the position of rational surface, $\rho \equiv r/a$ is the normalized minor radius.

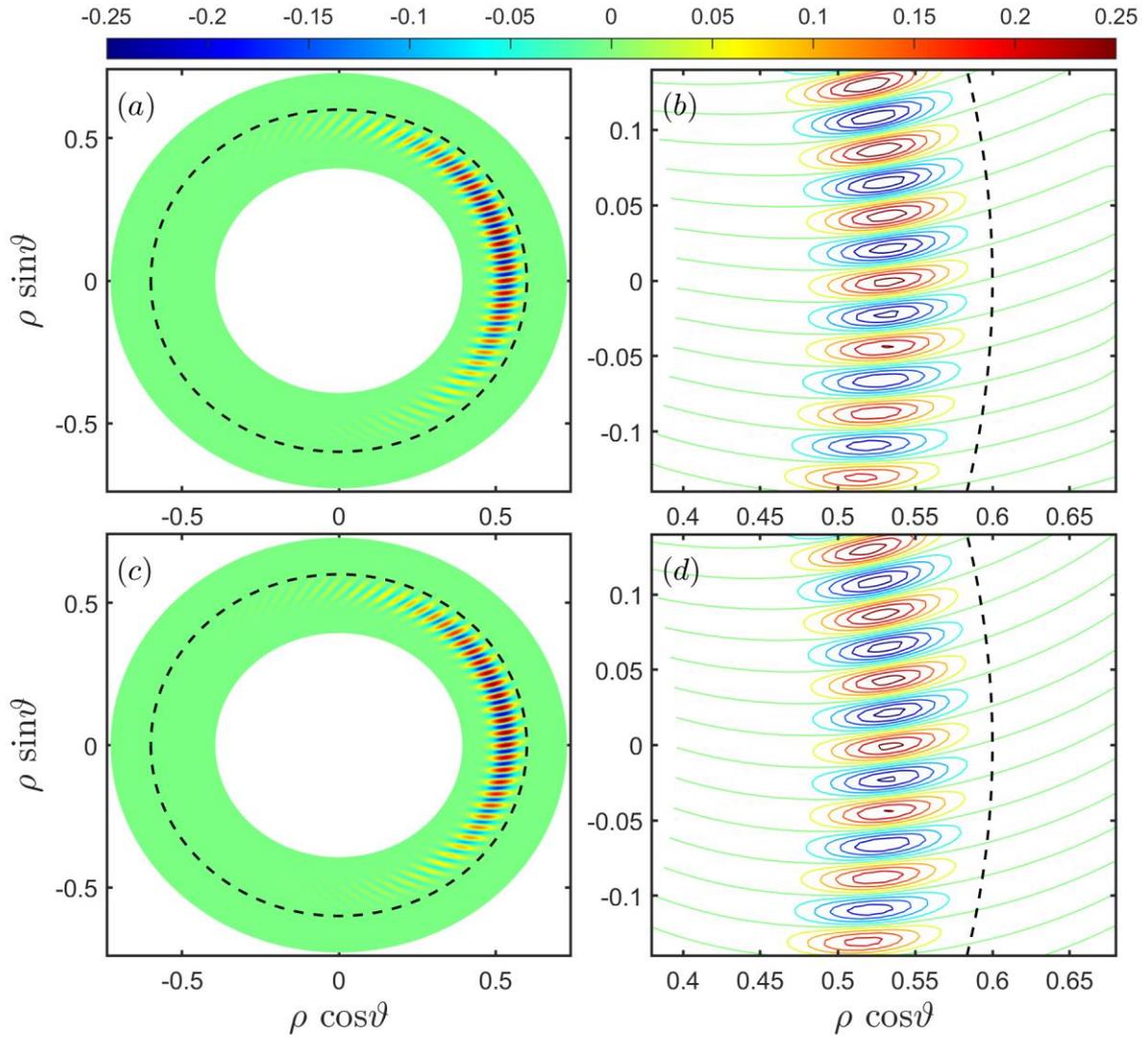

Figure 7B. Real part of the two dimensional wave function $\mathrm{Re}\varphi(r,\vartheta)$, $n = -50$, $\eta_i = 5.0$.

(a) WABT, (c) SIPM. (b/d): close-up in bad curvature regime of (a/c) respectively.

The black dashed line represents the position of rational surface, $\rho \equiv r/a$ is the normalized minor radius.

The poloidal width of the "mode B" is much narrower than "mode A" because of higher *n*. For B, there are 13 contours (Figure 7B (d)), but only 5 for A (Figure 7A (d)); the difference arises because high poloidal number implies reduced poloidal wavelength. The actual radial extension for B is quite smaller than A, even though they are similar when measured by *x*, which is proportional to *n*.

## 6. Warm ion branch - physics behind eigenmode crossing zero frequency

The observation of drift wave eigenmode crossing zero frequency is *not* new. It has been reported in the numerical solution of the ballooning equation a few decades ago, *e.g.*, Fig.2 of [37] and Fig.4 of [38]. It is interesting that this rather spectacular phenomenon did not draw the community attention in general and particularly in the context of the origin of edge turbulence propagating in the electron direction [1-5]. It is, perhaps, time to search for and understand the physics behind propagation reversal. In the process, we will also attempt to probe its relationship, if any, with the well-known drift wave CIED propagating, solely, in electron diamagnetic direction.

Since the warm ions have nothing to do with the curvature term, a slab model analysis should be sufficient to reveal two branches of the mode. The slab limit of the eigenmode equation is simply the first term of Equation (6) for $l = 0$. It is a Weber equation that can be solved analytically. The quadratic dispersion relation has two roots obtained by invoking the well-known outgoing wave boundary condition [22], namely $\hat{\omega}_+$ and $\hat{\omega}_-$, as defined in Equation (D6). A simple analysis (see Appendix D) illustrates that $\hat{\omega}_+$ corresponds to cold ion branch of the drift wave, which in cold ion limit approaches $\omega_+ \to \omega_{*e}(1 - iL_n/L_s)/(1 + \hat{k}_\vartheta^2)$, with explicit shear damping (that stabilizes it against any dissipative driving). The $\hat{\omega}_-$ corresponds to the warm ion branch that is proportional to $\tau_i$. In cold ion limit, it is zero. The combined dispersion for these two modes shows two independent modes coupled through $\tau_i$; it is of the generic form $\omega(\omega - \omega_{*e} \dots) = -i\tau_i \dots$. Noticeably, the "$i$" in the dispersion is brought into the non-dissipative system by the outgoing wave boundary condition. It has the same origin - the right hand side of Equation (D4) that stabilizes $\hat{\omega}_+$, but destabilizes $\hat{\omega}_-$ (also known as 'reactive instability' in literature). To the first order of $L_n/L_s \ll 1$ (the lowest order of finite magnetic shear) it is a purely growing mode driven by magnetic shear, which is consistent with the observation of [39] and [40]. The real frequency appears from the second order of $L_n/L_s$ as shown in Equation (D9). The computation of the warm ion branch $\hat{\omega}_-$ in the slab model is displayed in Figure D1 and D2 for relevant parameters. It is always unstable and is in the ion diamagnetic direction.

Notice that, the same term in Equation (D4) plays opposite role for the different branches. This is reminiscent of the concept of positive *vs.* negative energy waves [41-44]. In terms of this nomenclature, one may say that CIED is a positive energy wave, whereupon shear damping is natural (shear cut-off the mode tail in radial direction induces a damping force), and WID is a negative energy wave, driven by magnetic shear. The sign of wave energy is determined here by what follows from the more physical O'Neil criterion stated in the first paragraph of [41] - "A plasma wave is said to have positive energy if energy must be added to the plasma when the wave is excited. Likewise, a wave is said to have negative energy if energy must be removed from the plasma when the wave is excited" - without actually calculating the sign of wave energy $\omega(\partial\varepsilon/\partial\omega)$ for an electrostatic wave like in [42-44]. Thus, for our non-dissipative system the drift wave in warm (cold) ion branch $\hat{\omega}_-$ ($\hat{\omega}_+$) may be identified as the negative (positive) energy wave.

Inclusion of curvature effects modifies the dispersion of the waves in the generic drift fluid model; it can be cast into the form $(\omega - \hat{\omega}_+)(\omega - \hat{\omega}_-) = \hat{\kappa} \dots$ in small $\hat{\kappa}$ limit, where $\hat{\kappa}$ stands for the normalized curvature. This dispersion could be analytically continued to the fully toroidal solution ($\hat{\kappa}: 0 \to 1$). The analytic continuation is shown in sub-section 6.1. The sub-section 6.2 is devoted to discussing the WID threshold.

## 6.1. Magnetic Curvature induced Coupling of negative and positive energy waves

It is straightforward to perform the coupling by mapping $\hat{\kappa}: 0 \to 1$ via running 1D-BT starting from the unstable branch $\omega = \hat{\omega}_-$ as if $\hat{\kappa}$ were the working parameter in front of the curvature term of Equation (14) [the second term of Equation (15)]. The results are displayed for $n = -20, -50$ and $\eta_i = 1.0, 5.0$, as shown in Figures 8A and 8B. It is remarkable to observe that the coupling augments the growth rate by a factor ~3. The physics behind the growth rate augmentation due to curvature coupling is quite different from that for CIED mode. In the latter case, curvature coupling simply acts to reduce shear damping; it is not a direct driving force. In contradistinction, the coupling between negative and positive energy wave results in a strong driving force to reactive instability [41]. Then the coupling of the negative energy wave to positive energy wave via magnetic curvature provides the major driving force for WID.

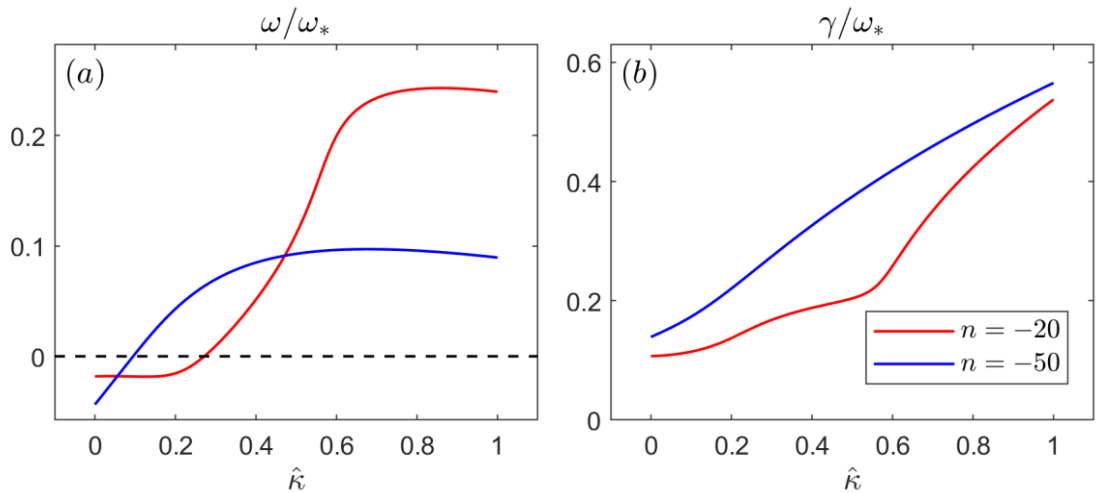

Figure 8A. Normalized $\hat{\omega}_-$ ($\hat{\kappa}: 0 \to 1$) for $\eta_i = 1.0$

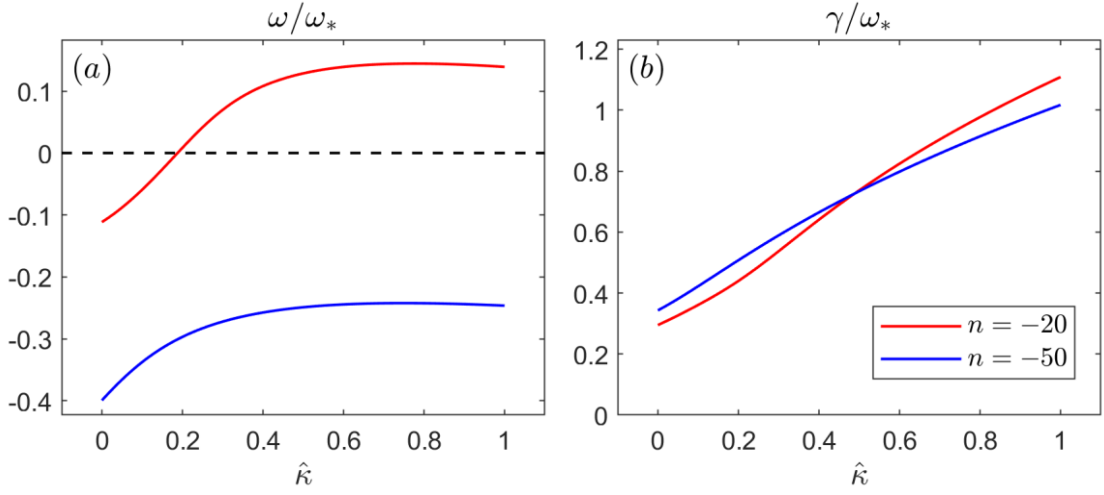

Figure 8B. Normalized $\widehat{\omega}_-$ ($\hat{\kappa}: 0 \to 1$) for $\eta_i = 5.0$

One may directly check that the results are consistent with 1D-BT curves - the dashed line of Figure 2. The regularity of the mapping shown in Figure 8A and 8B are double checked by the inverse mapping ($\hat{\kappa}: 1 \to 0$). The consistency between the two confirms the claim that the phenomenon of the eigenmode crossing zero frequency originates with the negative energy wave $\widehat{\omega}_-$, as analytic continuation via curvature coupling to the positive energy wave $\widehat{\omega}_+$. One may also notice that such a coupling is destabilizing since $\text{Im}[\widehat{\omega}(\hat{\kappa})]$ increases monotonically with the coupling strength.

For most parameters the coupling to positive energy wave $\widehat{\omega}_+$ brings a drastic change in real frequency; it changes from the ion to electron direction except for relatively large $\eta_i = 5.0$. In the displayed case, the real frequency $\text{Re}[\widehat{\omega}_-(0)] \approx -0.4$; it is almost as large as the growth rate. Unlike the other cases where the coupling to positive energy wave alters the imaginary potential function quite a bit (from well to hump) in Figure 9B, this is the exception as displayed for $n = -50$, $\eta_i = 5.0$, in Figure 9A, where the well structure remains in global scale as shown in Figure 9A.

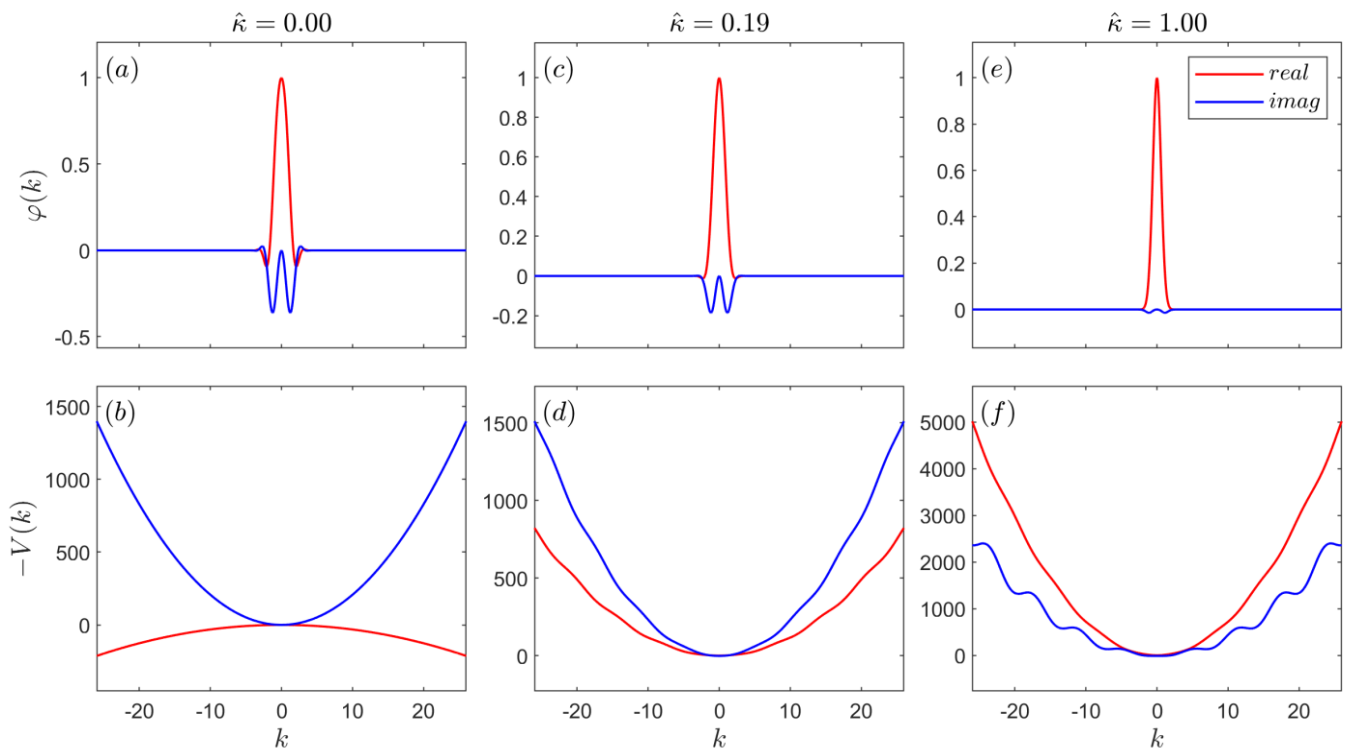

Figure 9A. Potential and the mode structure under the mapping for $n = -50$, $\eta_i = 5.0$

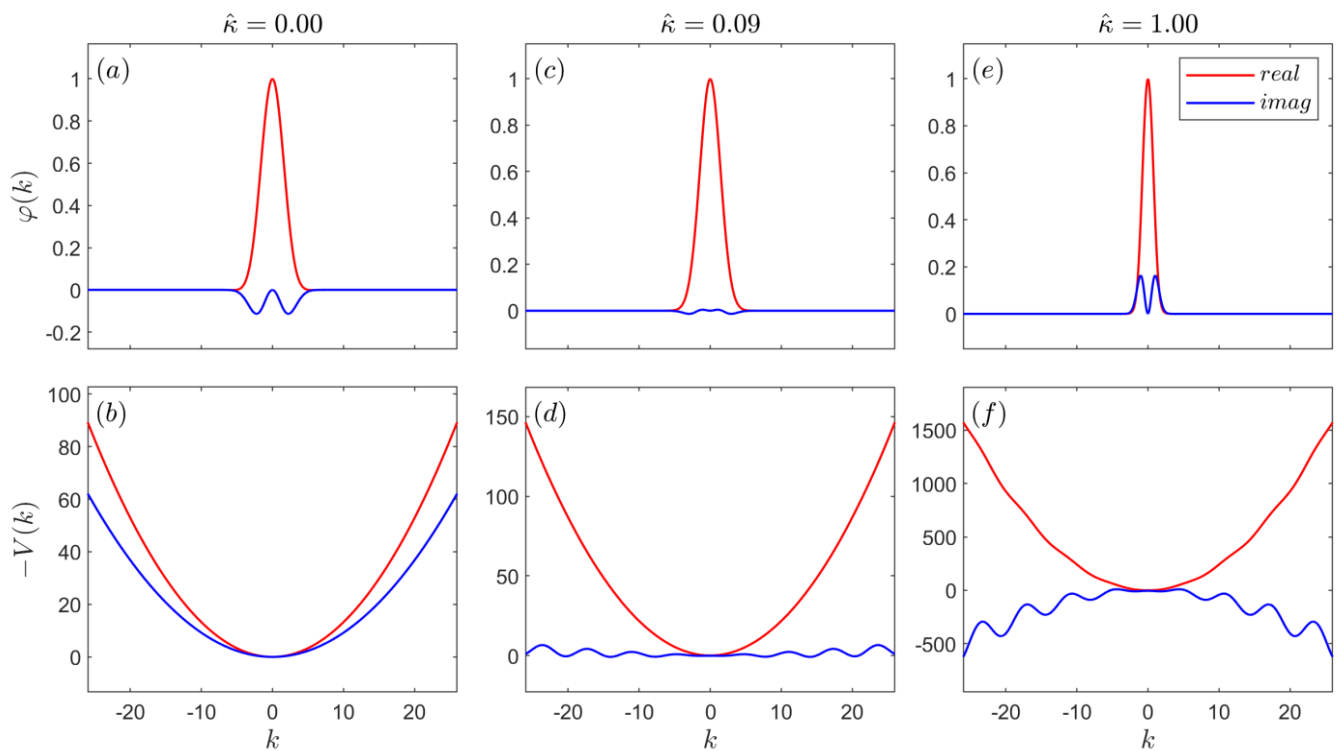

Figure 9B. Potential and the mode structure under the mapping for $n = -50$, $\eta_i = 1.0$

### 6.2. Threshold of WID mode

For lower ion temperature WID mode is getting stabilized. The eigenvalue is illustrated in Figure 10, where the solid (dotted) line stands for the result from 1D-BT (WABT) scanning over $\bar{\eta}_i$ for $n = -30,\ -40,\ -50$. While $\eta_i$ and $\tau_i$ are combined into one parameter $\bar{\eta}_i \equiv \tau_i(1+\eta_i)$ in 1D-BT, they are somewhat separated in WABT arising from TSB terms. The actual scan for dotted curves goes over $\tau_i$ for fixed $\eta_i = 2.0$. The 2D effect becomes unimportant as approaching to WID threshold for growth rate. As demonstrated in Figure 10, the term 'threshold of ITG mode' makes little sense, since WID is marginally stable only when the propagation is in the electron direction. It must be emphasized, however, that there is a wide range of tokamak parameters where WID mode is unstable. One could, perhaps, view WID wave as the modern-day version of the 'universal drift wave'.

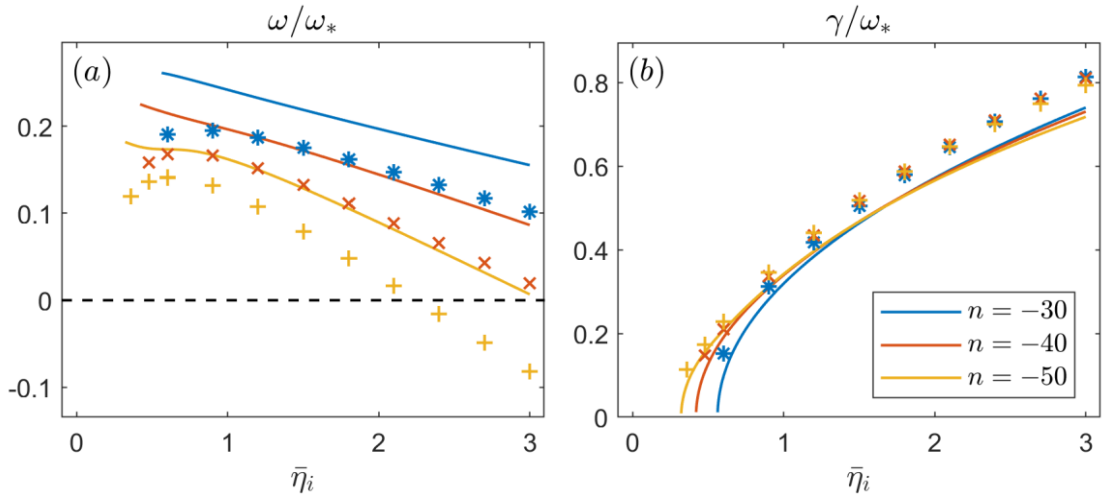

Figure 10. The eigenvalue of WID mode approaching marginal stability for 1D-BT (solid line) and WABT (dotted line) versus $\bar{\eta}_i$ for $n = -30,\ -40,\ -50$. The scan for WABT goes over $\tau_i$ for fixed $\eta_i = 2.0$.

## 7. Summary and conclusion

The WID wave eigenmode in the generic drift fluid model is thoroughly studied from sheared slab to a toroidal geometry. In the slab limit it belongs to the warm ion branch, $\hat{\omega}_- \approx i\mathbb{S}\bar{\eta}_i - \mathbb{S}^2\bar{\eta}_i(1+\bar{\eta}_i) + O(\mathbb{S}^3)$, a purely growing mode to the lowest order of $\mathbb{S}$, driven by the magnetic shear and propagating in ion direction. In the cold ion limit, it vanishes. It is the negative energy wave per the O'Neil criterion [41], while the other solution of the dispersion is the well-known CIED wave, stabilized by the same magnetic shear $\omega_+ \approx \omega_{*e}(1-i\mathbb{S})$; CIED is the positive energy wave also per the O'Neil criterion [41]. In a torus, the two modes are coupled via magnetic curvature, resulting in significant changes in eigenvalue. In general, the toroidal coupling boosts up the growth rate by a factor ~3. Therefore, the major driving force of WID mode arises from the coupling between the negative energy wave to positive energy wave via magnetic curvature. The real frequency change can be so large that the propagation can change from the ion to the electron diamagnetic direction. Such a tendency becomes stronger for lower $n$ and $\eta_i$. On the opposite end (higher $n$ and $\eta_i$) the real frequency may remain in the ion direction (Figure 8B). In a broader parameter domain, relevant to experiments near tokamak edge, the WID wave is either in the electron direction or near zero frequency.

One must point out here that the community has not paid much attention to the warm ion branch present in the slab model; it is missing, for example, in two rather influential review articles [45-46]. That, perhaps, explains why the physics behind the ITG to WIED transition has not been uncovered. With our understanding of WID mode a term like 'threshold of ITG' (at some $\eta_{i,\text{crit}}$) makes little sense; ITG is never stabilized, it transits smoothly to WIED with lower $\eta_i$. Since fusion tokamak is always in warm ion state, WID wave is expected to be always unstable. Therefore, as stated earlier, one could view WID as the modern-day version of the 'universal drift wave'.

Our search for the full solution of the system begins with 1D-BT. While the ballooning equation contains two branches with the natural boundary condition, one is apt to see only the WID branch in shooting code because it is unstable in the non-dissipative model. The analytic continuation ($\hat{\kappa}: 0 \to 1$) is shown to be regular on the upper half complex frequency plane. Beyond 1D-BT, the eigenvalue problem is solved by making use of the 2D ballooning theory WABT, where the 2D natural boundary value problem is decomposed into two 1D natural boundary value problems. The WABT solution is further refined by the numerical approach SIPM. Both eigenvalue and 2D mode structure of WABT and SIPM are displayed for a single mode number $n$ at the specific rational surface $r_j$, known as the central rational surface of the *local* toroidal mode. *Inter alia*, two parameters $\eta_i \equiv L_n/L_{T_i}$ and $\rho_i k_\vartheta$ are more relevant to the transition between directions of propagation; the lower (higher) $\eta_i$ and $\rho_i k_\vartheta$ favors the electron (ion) direction.

The physics behind WID mode could also be related to the frequency reversal for TEM (as noticed by Jarmén *et. al.* [38]) and ubiquitous mode [47, 48]. In perspective, the WIED mode may also shed light on the origin of trapped-particle-free edge turbulence: (1) observed on TEXT in early 1990's [1-4] in L-mode discharge, and (2) observed in various coherent modes like the I-mode [9-16]. In addition, it is likely to contribute to the edge localized mode (ELM) breaking the operation regime of the H-mode; the typical ELM toroidal number is expected to be lower than 20.

**Appendix A. Derivation of the generic drift fluid model in $(x, l)$ representation**

The 4 operators in Equation (4) in $(x, l)$ representation are

$$\nabla_\perp^2 \hat{\varphi} \approx \left( \frac{\partial^2}{\partial r^2} + \frac{1}{r^2} \frac{\partial^2}{\partial \vartheta^2} \right) \hat{\varphi} \rightarrow k_\vartheta^2 \hat{s}^2 \frac{d^2 \varphi_l}{dx^2} - \frac{(m+l)^2}{r^2} \varphi_l, \tag{A1}$$

$$\nabla_\parallel \hat{\varphi} = \frac{1}{qR} \left( \frac{\partial}{\partial \vartheta} + q \frac{\partial}{\partial \zeta} \right) \hat{\varphi} \rightarrow \frac{i}{qR}(x-l)\varphi_l, \tag{A2}$$

$$\mathbf{v}_* \cdot \nabla \hat{\varphi} = -\rho_s c_s \frac{\partial \ln n_{i0}}{\partial r} \frac{\partial \hat{\varphi}}{r \partial \vartheta} \rightarrow i\rho_s c_s \frac{\partial \ln n_{i0}}{\partial r} \frac{m+l}{r} \varphi_l, \tag{A3}$$

$$\mathbf{v}_D \cdot \nabla \hat{\varphi} = \frac{2\rho_s c_s}{R} \left( \sin\vartheta \frac{\partial}{\partial r} + \cos\vartheta \frac{\partial}{r \partial \vartheta} \right) \hat{\varphi} \rightarrow -i \frac{\rho_s c_s}{R} \left[ k_\vartheta \hat{s} \frac{d}{dx}(\varphi_{l+1} - \varphi_{l-1}) + \frac{m+l}{r}(\varphi_{l+1} + \varphi_{l-1}) \right]. \tag{A4}$$

The variations of $R$ and $B$ are neglected in this paper, $R \approx R_0$, $B \approx B_0$. Use is made of the linear equilibrium profile

$$r = r_j \left( 1 + \frac{x}{m\hat{s}} \right) \equiv r_j F_r(x), \tag{A5}$$

$$q = q_j \left( 1 + \frac{x}{m} \right) \equiv q_j F_q(x), \tag{A6}$$

$$n_{i0} = n_{0,j} \left( 1 - \frac{r_j}{L_n} \frac{x}{m\hat{s}} \right) \equiv n_{0,j} F_n(x), \tag{A7}$$

$$T_{s0} = T_{s0,j} \left( 1 - \frac{r_j}{L_{T_s}} \frac{x}{m\hat{s}} \right) \equiv T_{s0,j} F_{T_s}(x) \quad (s=i,e). \tag{A8}$$

Substituting (A1-A8) into Equation (4), we get the eigenmode equation in $(x, l)$ representation

$$\hat{k}_\vartheta^2 \hat{s}^2 \frac{d^2 \varphi_l}{dx^2} - \hat{k}_\vartheta^2 \frac{1}{F_r^2}(1+l/m)^2 \varphi_l + (x-l)^2 \frac{1}{F_q^2} \frac{\hat{\omega}_s^2}{\hat{\omega}^2} \varphi_l$$
$$- \frac{1}{F_{T_e}} \frac{\hat{\omega} - (F_{T_e}/F_r F_n)(1+l/m)}{\hat{\omega} + \tau_{i,j}(F_{T_i}/F_r F_n)(1+\eta_{i,j} F_n/F_{T_i})(1+l/m)} \varphi_l \tag{A9}$$
$$- \frac{\hat{\omega}_D}{2\hat{\omega}} \left[ \hat{s} \frac{d}{dx}(\varphi_{l+1} - \varphi_{l-1}) + \frac{1}{F_r}(1+l/m)(\varphi_{l+1} + \varphi_{l-1}) \right] = 0$$

where $\omega$ and other parameters are normalized as in Section 2.

Equation (A9) is essentially a 2D eigenvalue problem, however, in the approximation of large $m$, $l/m \ll 1$, $x/m \ll 1$, it reduces to a 1D eigenvalue problem by introducing the concept of translational invariance. The equation is unchanged under the transformation $(x, l) \rightarrow (x+1, l+1)$, implying translational symmetry in the system [22]. The sideband $l/m$ and slow variation of equilibrium profile $x/m$ violate the translational invariance, and are all called translational symmetry broken (TSB) terms. For large $m$, translational invariance is the leading term, TSB terms are small enough to be retained up to second order $O(l^2/m^2)$, with the approximation $F(x) \approx F(l)$. Then the generic drift fluid model expanded to include TSB terms in $(x, l)$ representation is obtained as Equation (6), where

$$A_1(\hat{\omega}) = -2\frac{\hat{s}-1}{\hat{s}}\hat{k}_g^2 + \frac{(\hat{\omega}-1)}{\hat{s}(\hat{\omega}+\bar{\eta}_{i,j})^2}\left[(\hat{s}-1)\bar{\eta}_{i,j} + \tau_{i,j}\left(\frac{r_j}{L_n} - \frac{r_j}{L_{T_i}}\right)\right]$$
$$+ \frac{1}{\hat{s}(\hat{\omega}+\bar{\eta}_{i,j})}\left(\hat{s}-1+\frac{r_j}{L_n} - \hat{\omega}\frac{r_j}{L_{T_e}}\right) \tag{A10}$$

and

$$A_2(\hat{\omega}) = -\frac{(3-4\hat{s}+\hat{s}^2)}{\hat{s}^2}\hat{k}_g^2 + \frac{1}{\hat{s}^2(\hat{\omega}+\bar{\eta}_{i,j})}\left[(\hat{s}-1)\left(\frac{r_j}{L_n}-1\right) + \frac{r_j^2}{L_n^2} - \hat{\omega}\frac{r_j^2}{L_{T_e}^2}\right]$$
$$-\frac{1}{\hat{s}^2(\hat{\omega}+\bar{\eta}_{i,j})^2}\left[(\hat{s}-1)\bar{\eta}_{i,j} + \tau_{i,j}\left(\frac{r_j}{L_n} - \frac{r_j}{L_{T_i}}\right)\right]\left(\hat{s}-1+\frac{r_j}{L_n} - \hat{\omega}\frac{r_j}{L_{T_e}}\right)$$
$$-\frac{\tau_{i,j}(\hat{\omega}-1)}{\hat{s}^2(\hat{\omega}+\bar{\eta}_{i,j})^3}\left\{-(\hat{\omega}+\bar{\eta}_{i,j}-\tau_{i,j})\frac{r_j^2}{L_n^2} + \left[-(\hat{s}-1)(\hat{\omega}-\bar{\eta}_{i,j}) + (\hat{\omega}+\bar{\eta}_{i,j}-2\tau_{i,j})\frac{r_j}{L_{T_i}}\right]\frac{r_j}{L_n}\right. \tag{A11}$$
$$\left. + \left[(\hat{s}-1)(\hat{\omega}+\hat{s}\bar{\eta}_{i,j})\frac{\bar{\eta}_{i,j}}{\tau_{i,j}} + (\hat{s}-1)(\hat{\omega}-\bar{\eta}_{i,j})\frac{r_j}{L_{T_i}} + \tau_{i,j}\frac{r_j^2}{L_{T_i}^2}\right]\right\}$$

**Appendix B. 2D shooting code for solution of WABT**

Starting with Equation (18) and introducing the transform [11]

$$\psi(\lambda) = \Phi(\lambda)\exp\left[-\frac{1}{2}\int^\lambda d\lambda' P(\lambda')\right]. \tag{B1}$$

In the large $n$ limit, the equation for $\Phi(\lambda)$ is

$$\frac{d^2\Phi}{d\lambda^2} + \frac{n^2}{\bar{L}_2^{(0)}(\lambda;\hat{\omega})}\left\{\Omega(\lambda) - \Omega(\hat{\omega}) + \frac{[\bar{L}_1^{(0)}(\lambda;\hat{\omega})]^2}{4\bar{L}_2^{(0)}(\lambda;\hat{\omega})}\right\}\Phi = 0. \tag{B2}$$

The steps in the iterative procedure [12] are listed below:

(I) Begin with an initial guess (solution of 1D-BT), $\hat{\omega} \to \hat{\omega}^{(0)}$ to solve Equation (17)

$$\left[L_0(k,\lambda;\hat{\omega}^{(0)}) - \Omega^{(0)}(\lambda)\right]\chi^{(0)}(k,\lambda) = 0, \tag{B3}$$

by imposing the boundary condition, Equation (16) for all $\lambda$.

(II) Substitute $\chi^{(0)}(k,\lambda)$ into Equation (21) to compute $\bar{L}_1^{(0)}(\lambda;\hat{\omega}^{(0)})$ and $\bar{L}_2^{(0)}(\lambda;\hat{\omega}^{(0)})$, and consequently obtain the equation for $\Phi(\lambda)$

$$\frac{d^2\Phi^{(0)}(\lambda)}{d\lambda^2} + \frac{n^2}{\bar{L}_2^{(0)}(\lambda;\hat{\omega}^{(0)})} \left\{ \Omega^{(0)}(\lambda) - \Omega(\hat{\omega}^{(1)}) + \frac{\left[\bar{L}_1^{(0)}(\lambda;\hat{\omega}^{(0)})\right]^2}{4\bar{L}_2^{(0)}(\lambda;\hat{\omega}^{(0)})} \right\} \Phi^{(0)}(\lambda) = 0. \tag{B4}$$

This equation is solved with evanescent boundary conditions.

(III) The global eigenvalue $\hat{\omega}^{(1)}$ follows from Equation (9) by substituting $\Omega(\hat{\omega}^{(1)})$, the eigenvalue of Equation (B4). Repeat the procedures (I-III) to obtain $\hat{\omega}^{(i+1)}$ from $\hat{\omega}^{(i)}$ until $|1 - \hat{\omega}^{(i+1)}/\hat{\omega}^{(i)}| < \varepsilon$ with $\varepsilon = 10^{-4}$ as the convergence condition. The convergence was usually achieved after less than 10 iterations.

## Appendix C. Iterative finite difference method of SIPM

The 2D eigenvalue problem Equation (6) is put in the form as $T[\partial/\partial x, l; \hat{\omega}]\varphi_l(x) = \Omega(\hat{\omega})\varphi_l(x)$, where $T$ is a differential operator with derivative of $x$, it also depends on $l$ and the eigenvalue $\hat{\omega}$. The spatial discrete grids are $x_k = k \cdot h$ $(k = -K, -K+1, \ldots, K-1, K)$, the step size is $h = (x_r - x_l)/2K$, $x_l$ ($x_r$) is the left (right) boundary. The $l$ grids are $l = -L, -L+1, \ldots, L-1, L$, and $\varphi_l(x)$ is cut-off at large $|l| > L$. Use is made of the central difference for derivative of $x$ to yield the matrix equation

$$\mathbf{M}(\hat{\omega}) \cdot \mathbf{\Phi} = \Omega(\hat{\omega})\mathbf{\Phi}, \tag{C1}$$

where $\mathbf{\Phi} = (\mathbf{\Phi}_{-L}, \mathbf{\Phi}_{-L+1}, \cdots, \mathbf{\Phi}_{L-1}, \mathbf{\Phi}_L)^T$, $\mathbf{\Phi}_l = (\varphi_{l,-K}, \varphi_{l,-K+1}, \cdots, \varphi_{l,K-1}, \varphi_{l,K})^T$, $\varphi_{l,k} \equiv \varphi_l(x_k)$. $\mathbf{M}$ is block tri-diagonal matrix and its dimension is $(2L+1)(2K+1) \times (2L+1)(2K+1)$.

The eigenvalue problem, Equation (C1) can only be solved by an iterative method because it is nonlinear in $\hat{\omega}$. The iterative procedure is listed below:

(I) It begins with the eigenvalue of WABT solution as initial guess, $\hat{\omega}_{\text{WABT}} \to \hat{\omega}^{(0)}$ to compute the matrix $\mathbf{M}^{(0)} = \mathbf{M}(\hat{\omega}^{(0)})$, and eigenvalue $\Omega^{(0)} = \Omega(\hat{\omega}^{(0)})$ via Equation (9).

(II) Wave functions $\varphi_l(x)$ of WABT solution at discrete grids $\mathbf{\Phi}^{(0)}$ are normalized, $\mathbf{\Phi}^{(0)} = \mathbf{\Phi}^{(0)}/\|\mathbf{\Phi}^{(0)}\|_2$, the matrix equation $(\mathbf{M}^{(0)} - \Omega^{(0)}) \cdot \mathbf{\Phi}^{(1)} = \mathbf{\Phi}^{(0)}$ is solved for $\mathbf{\Phi}^{(1)}$ by LU decomposition.

(III) Compute $\Delta\Omega^{(1)} = (\mathbf{\Phi}^{(1)})^\dagger \cdot (\mathbf{M}^{(0)} - \Omega^{(0)}) \cdot \mathbf{\Phi}^{(1)} / (\mathbf{\Phi}^{(1)})^\dagger \cdot \mathbf{\Phi}^{(1)}$ and go back to step (II), $\mathbf{\Phi}^{(0)} \to \mathbf{\Phi}^{(1)}$ until $\delta = \|\Delta\Omega^{(1)} \cdot \mathbf{\Phi}^{(1)} - \mathbf{\Phi}^{(0)}\|_\infty < 10^{-6}$.

(IV) Substitute the new eigenvalue $\Omega^{(1)} = \Omega^{(0)} + \Delta\Omega^{(1)}$ into Equation (9) to obtain the global eigenvalue $\hat{\omega}^{(1)}$.

In the preceding lines, $\mathbf{\Phi}^\dagger$ stands for the Hermitian conjugate of $\mathbf{\Phi}$, $\|\mathbf{\Phi}\|_2 \equiv \sqrt{\sum_i \Phi_i^2}$ and $\|\mathbf{\Phi}\|_\infty \equiv \max_i |\Phi_i|$ are the 2-norm and $\infty$-norm of vector $\mathbf{\Phi}$ respectively.

(V) Repeat the steps (I-IV) to obtain $\hat{\omega}^{(i+1)}$ from $\hat{\omega}^{(i)}$ until $\left|1 - \hat{\omega}^{(i+1)}/\hat{\omega}^{(i)}\right| < \varepsilon$ with $\varepsilon = 10^{-4}$ as the convergence condition.

The convergence was usually achieved after less than 10 iterations. The relative difference of the eigenvalue solution of SIPM and WABT, are usually much smaller than $1/n$, implying good agreement.

## Appendix D. Two branches of the eigenmode in slab model

The eigenmode equation of Equation (4) in 1D slab model can be obtained from Equation (6) by setting $l$ to zero

$$\left[\hat{k}_g^2 \hat{s}^2 \frac{d^2}{dx^2} - \hat{k}_g^2 + \frac{\hat{\omega}_s^2}{\hat{\omega}^2} x^2 - \frac{\hat{\omega}-1}{\hat{\omega}+\bar{\eta}_i}\right]\hat{\varphi}_0 = 0. \tag{D1}$$

For convenience, introducing a new variable $\hat{x} \equiv (r - r_0)/\rho_s$, then Equation (D1) reduces to

$$\left[\frac{d^2}{d\hat{x}^2} + \frac{\mathbb{S}^2}{\hat{\omega}^2}\hat{x}^2 + \bar{\Omega}\right]\hat{\varphi}_0 = 0, \quad \bar{\Omega} \equiv \frac{1-\hat{\omega}}{\hat{\omega}+\bar{\eta}_i} - \hat{k}_g^2, \tag{D2}$$

where $\mathbb{S} \equiv L_n/L_s$ and $L_s \equiv \hat{s}/qR$. Equation (D2) is consistent with Eq. (7) of Ref. [40] without flow. The solution of Equation (D2) satisfying the outgoing wave boundary condition [22] is

$$\hat{\varphi}_0 = \exp\left[-i\frac{\mathbb{S}}{2\hat{\omega}}\hat{x}^2\right]. \tag{D3}$$

It yields the dispersion relation

$$\bar{\Omega} = i\frac{\mathbb{S}}{\hat{\omega}}. \tag{D4}$$

Explicitly,

$$\left(1+\hat{k}_g^2\right)\hat{\omega}^2 - \left(1-\hat{k}_g^2\bar{\eta}_i - i\mathbb{S}\right)\hat{\omega} + i\mathbb{S}\bar{\eta}_i = 0. \tag{D5}$$

Two solutions of the dispersion are

$$\hat{\omega}_\pm = \frac{1}{2\left(1+\hat{k}_g^2\right)}\left\{\left(1-\hat{k}_g^2\bar{\eta}_i - i\mathbb{S}\right) \pm \sqrt{\left(1-\hat{k}_g^2\bar{\eta}_i - i\mathbb{S}\right)^2 - 4i\mathbb{S}\bar{\eta}_i\left(1+\hat{k}_g^2\right)}\right\}. \tag{D6}$$

The square root of Equation (D6) can be expanded for small $\mathbb{S} \ll 1$, yielding

$$\hat{\omega}_\pm \approx \frac{1}{2\left(1+\hat{k}_g^2\right)}\left\{\left(1-\hat{k}_g^2\bar{\eta}_i - i\mathbb{S}\right) \pm \left(1-\hat{k}_g^2\bar{\eta}_i - i\mathbb{S}\right)\left[1 - \frac{2i\mathbb{S}\bar{\eta}_i\left(1+\hat{k}_g^2\right)}{\left(1-\hat{k}_g^2\bar{\eta}_i - i\mathbb{S}\right)^2}\right]\right\}. \tag{D7}$$

The $\hat{\omega}_+$ branch is the cold ion branch, surviving in $\bar{\eta}_i \to 0$ limit

$$\hat{\omega}_+ \approx \frac{1-i\mathbb{S}}{1+\hat{k}_g^2}. \tag{D8}$$

The $\hat{\omega}_-$ branch is the warm ion branch since it is proportional to the warm ion temperature

$$\hat{\omega}_- = \frac{i\mathcal{S}\bar{\eta}_i\left(1-\hat{k}_\vartheta^2\bar{\eta}_i\right)}{\left(1-\hat{k}_\vartheta^2\bar{\eta}_i\right)^2} - \frac{\mathcal{S}^2\bar{\eta}_i\left(1+\bar{\eta}_i\right)}{\left(1-\hat{k}_\vartheta^2\bar{\eta}_i\right)^3} + O\left(\mathcal{S}^3\right). \tag{D9}$$

To the lowest order of $\mathcal{S}$ it is a purely growing mode driven by magnetic shear with warm ion. This is consistent with the conclusion of [39] and [40]. The real frequency is in the ion direction.

In cold ion limit this branch vanishes. Noticeably, near $\hat{k}_\vartheta^2\bar{\eta}_i \geq 1$ it would become a stable mode. However, this parameter domain is not quite accessible practically for fluid model where $\hat{k}_\vartheta^2 \ll 1$ is required. We may accept the concept that the warm ion branch is proportional to $\bar{\eta}_i$; to the first order of $\mathcal{S}$ it is a purely growing mode; the real frequency is at the order of $O\left(\mathcal{S}^2\right)$ in the ion diamagnetic frequency as shown in Equation (D9).

The numerical computation of Equation (D6) is shown in Figure D1 and D2. All parameters are the same as Section 4 except those explicitly given here. The contour for real frequency (Re$\hat{\omega}_-$) and growth rate (Im$\hat{\omega}_-$) is shown in the $\bar{\eta}_i$-$|\rho_i k_\vartheta|$ parameter space of Figure D1 (a), (b) respectively.

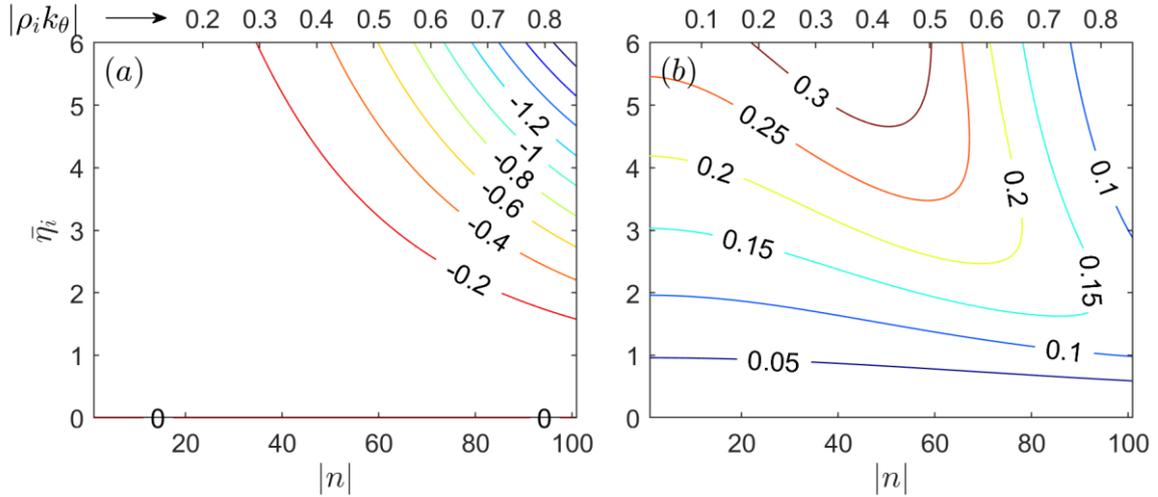

Figure D1. (a) The contour for real frequency Re$\hat{\omega}_-$ and (b) growth rate Im$\hat{\omega}_-$

The real frequency of $\hat{\omega}_-$ and growth rate versus $|\rho_i k_\vartheta|$ for four different $\bar{\eta}_i$ is shown in Figure D2. These results indicate that in the interested parameter domain the mode in warm ion branch is always in ion diamagnetic direction and unstable.

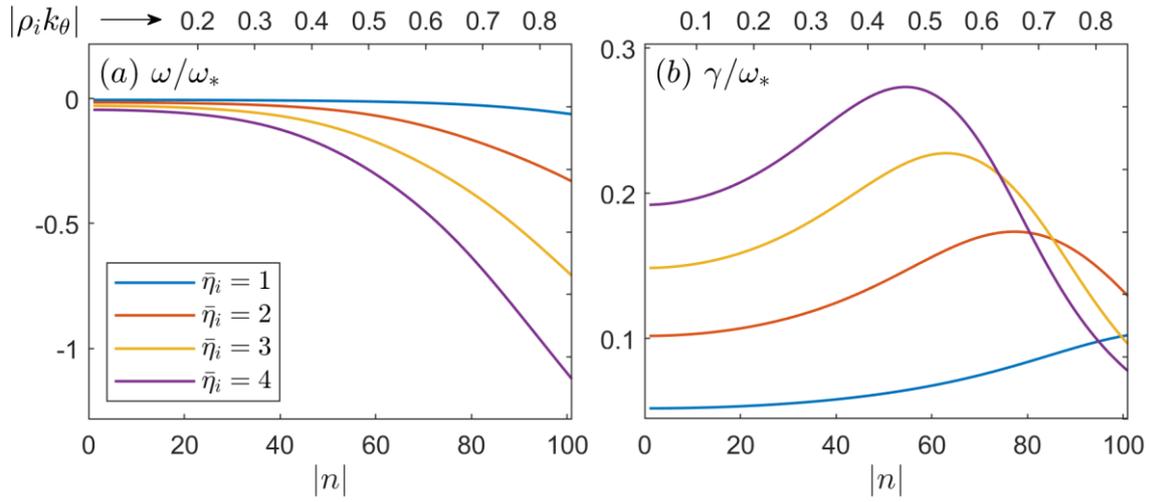

Figure D2. (a) real frequency of $\hat{\omega}_-$ and (b) growth rate versus $|\rho_i k_\vartheta|$ for 4 different $\bar{\eta}_i$

## Acknowledgment

One of the authors, Y. Z. Zhang acknowledges helpful discussion with Prof. J. Q. Dong. This work is supported by the National Natural Science Foundation of China (Grant Nos. U1967206, 11975231, 11805203, 11775222), the National MCF Energy R & D Program, China (Grant Nos. 2018YFE0311200, 2017YFE0301204), the Key Research Program of Frontier Sciences, Chinese Academy of Sciences (Grant No. QYZDB-SSW-SYS004), and the U. S. Dept. of Energy (No. DE -FG02-04ER-54742).